\documentclass[10pt,letterpaper]{article}
\pdfoutput=1
\usepackage{opex3}
\usepackage{color}
\usepackage{cite}
\usepackage{amsmath,amssymb,graphicx}
\usepackage{epstopdf}
\begin{document}

\title{Frequency combs and platicons in optical microresonators with normal GVD}

\author{V.E. Lobanov$^{1}$, G. Lihachev$^{1,2}$,  T. J. Kippenberg$^{3}$ and M.L. Gorodetsky$^{1,2,*}$}

\address{$^{1}$Russian Quantum Center, Skolkovo 143025, Russia\\
$^{2}$Faculty of Physics, M. V. Lomonosov Moscow State University, Moscow 119991, Russia\\
$^{3}$Ecole Polytechnique Federale de Lausanne, CH~1015, Lausanne, Switzerland}

\email{*mg@rqc.ru}



\begin{abstract*}
We predict the existence of a novel type of the flat-top dissipative solitonic pulses, ``platicons'', in microresonators with normal group velocity dispersion (GVD).  We propose methods to generate  these platicons from cw pump. Their duration may be altered significantly by tuning the pump frequency. The transformation of a discrete energy spectrum of dark solitons of the Lugiato-Lefever equation into a quasicontinuous spectrum of platicons is demonstrated. Generation of similar structures is also possible with bi-harmonic, phase/amplitude modulated pump or via laser injection locking.
\end{abstract*}


%Microcavity devices (Lasers and laser optics)
%Harmonic generation and mixing (Nonlinear optics)
%Nonlinear optics, four-wave mixing (Nonlinear optics)
%For a complete list of OCIS codes, visit: http://www.opticsinfobase.org/submit/ocis/

%%%%%%%%%%%%%%%%%%%%%%% References %%%%%%%%%%%%%%%%%%%%%%%%%
%% Do not include separate BibTeX files; if BibTeX is used,
%% paste the output (contents of .bbl file) here.

%\bibliographystyle{osajnl}
%\bibliography{platicon}

%%%%%%%%%%%%%%%%%%%%%%%%%%  body  %%%%%%%%%%%%%%%%%%%%%%%%%%\section{Introduction}
High-quality factor nonlinear optical whispering gallery mode and ring-type microresonators are attracting growing attention as a promising platform for optical frequency-comb generation \cite{DelHaye07,Savchenkov08,Levy10,Razzari10,Kippenberg11,Papp13,Li12}.

When a microresonator with anomalous group velocity dispersion is pumped by a c.w. laser, a frequency comb of equidistant optical lines in spectral domain can appear due to cascaded four-wave mixing (FWM). The spacing between the lines corresponds to the free spectral range (FSR, typically several gigahertz up to terahertz) of the fundamental azimuthal modes, which is the inverted round-trip time. Although it was shown that FWM-based microresonator combs can perform at the level required for optical-frequency metrology applications \cite{Papp13,DelHaye08,DelHaye12}, such systems often suffer from significant frequency and amplitude noise \cite{Ferdous11,Papp11,DelHaye11} due to the formation of sub-combs \cite{Herr12}. Unlike conventional mode-locked laser-based frequency combs, microresonator based combs generally do not correspond to stable ultrashort pulses in time domain because of arbitrary phase relations between the comb lines obtained in the process of formation. This obstacle is possible to overcome using soliton regime, when a c.w. laser beam is converted into a train of femtosecond bright soliton pulses, corresponding to a low-noise smooth spectral envelope frequency comb in spectral domain. This regime was demonstrated recently experimentally in optical crystalline and integrated ring microresonators \cite{Herr14,Brasch14}. Temporal solitons allow combined with external broadening to achieve counting the cycles of light i.e. achieve an microwave to optical link \cite{Jost14}.

Obtaining anomalous GVD in broad band for arbitrary centered wavelength is challenging in microresonators since material GVD in the visible and near IR is mostly normal. Therefore, elucidation of new strategies that enable implementation of frequency combs using materials with normal GVD is of a significant interest. Recently, frequency comb and pulse generation from normal GVD microresonators has been theoretically predicted \cite{Matsko12,Hansson13,Godey14}. At the same time narrow comb-like spectra from a normal GVD CaF$_2$ \cite{Coillet13} and MgF$_2$ \cite{Liang14, Liang14b} crystalline resonators were demonstrated experimentally. It was also shown via numerical simulation that microresonators with normal GVD may support dark optical solitons \cite{Godey14} presumably demonstrated in integrated ring SiN microresonators \cite{Xue14}. It is known, however, that mode spectra in real microresonators significantly deviate from theoretical ones due to mode coupling between different families of modes \cite{Herr14b}. Individual resonant normal frequencies of coupled modes are therefore shifted due to avoided mode crossing. Frequency relations between the pump mode and two symmetrical sideband modes, one of which is shifted, may, therefore, constitute equivalent local anomalous dispersion \cite{Savchenkov12,Liu14,Herr14b} supporting ''soft'' excitation of the comb. In this case modulation instability may easily occur, generating primary comb lines which can subsequently excite wider comb with bright or dark solitons.

In this paper we show through numerical simulations that if just the pump mode eigenfrequency is shifted either due to the mode coupling \cite{Herr14b} or alternatively as a result of strong self-injection locking \cite{Liang14}, it is possible to generate in microresonators with normal GVD stable ultrashort flat-top pulses, which we call platicons \footnote{Analogous nonstationary pulses in fibers were experimentally observed and analysed in \cite{Varlot13} and called flaticons. The only drawback of this last name is that it is already widely used as FlatIcons. That is why we choose quite similar name originating from plateaus which are observed both in time and frequency dependencies.}.  We reveal that if this shift is large enough (a few line-widths), platicons may be generated spontaneously when the laser frequency is tuned through the effective zero detuning point of a high-Q resonance. In this solitonic regime one may continuously change the duration of generated flat-topped pulses varying the pump detuning.

We found that generation of wide platicons in normal GVD microresonators is significantly more effective in terms of transformation of c.w. power into power of pulse train than generation of bright solitons in microresonators with the same absolute value anomalous GVD.

\section{Numerical Simulation}
Our model is based on the system of dimensionless coupled mode equations \cite{Herr12,Chembo10}:
\begin{equation}\label{coupled_modes_eq}
\frac{\partial a_\mu}{\partial \tau}=-[1+i\zeta_{\mu}]a_\mu+i\sum_{\mu^\prime,\mu^{\prime\prime}} a_{\mu^\prime}a_{\mu^{\prime\prime}}a_{\mu^\prime+\mu^{\prime\prime}-\mu}^*+\delta_{0\mu}f,
\end{equation}
where
$a_\mu=A_\mu\sqrt{2g/\kappa}e^{-i(\omega_\mu-\omega_p-\mu D_1)t},$
can be interpreted as the slowly varying amplitude of the comb modes close to the mode frequency $\omega_\mu$, normalized detunings from equidistant frequency comb grid $\zeta_\mu=2(\omega_\mu-\omega_p-\mu D_1)/\kappa$,  $\tau=\kappa t/2$ denotes the normalized time, $D_1/(2\pi)=1/T_R$ is the FSR, $T_R$ is the roundtrip time, $g=\hbar\omega_0^2cn_2/n_0^2V_\mathrm{eff}$ is the nonlinear coupling coefficient, $V_\mathrm{eff}$ is effective mode volume, $n_2$ is nonlinear refractive index, $\kappa=\omega_0/Q=\kappa_0+\kappa_\mathrm{ext}$ denotes the cavity decay rate as the sum of intrinsic decay rate $\kappa_0=\omega_0/Q_0$ and coupling rate to the waveguide $\kappa_\mathrm{ext}=\omega_0/Q_\mathrm{ext}$, $Q$ is total quality factor, $Q_0$ is intrinsic quality factor, $Q_\mathrm{ext}$ is coupling Q-factor, $\eta=\kappa_\mathrm{ext}/\kappa$ is the coupling efficiency ($\eta=1/2$  for critical coupling), $f=[8\eta g P_\mathrm{in}/\kappa^2 \hbar \omega_0]^{1/2}$, stands for the dimensionless pump amplitude. The value $f=1$ corresponds to bistability and comb threshold \cite{Herr14}. All mode numbers $\mu$ are defined relative to the pumped mode $\mu=m-m_0$ with initial azimuthal number $m_0\simeq 2\pi R n_0/\lambda$ ($R$ -- the radius of the resonator, $\lambda$ -- the wavelength, $n_0$ -- the refractive index).

In order to take pump mode shift into account we modify the conventional dispersion law, neglecting third and higher-order dispersion: $\omega_\mu=\tilde\omega_0-\delta_{0\mu}\Delta+D_1\mu+\frac{1}{2}D_2\mu^2$,  where $\tilde\omega_0$ is the unperturbed  frequency of the pumped mode, $\Delta$ is the eigenfrequency shift and $\delta_{0\mu}$ is Kronecker's delta. For normal GVD $D_2<0$.

In numerical analysis we consider the following parameters, corresponding to crystalline MgF$_2$ microresonator: wavelength $\lambda=2\pi c/\omega_0=1.5$ $\mu$m, $n_0=1.37$, $n_2=0.9\times 10^{-20}$ m$^2$/W, $V_{\mathrm{eff}}=5.6\times 10^{-13}$m$^3$,  $Q_0=Q_\text{ext}=4\times 10^8$, $D_2/(2\pi)=-10$kHz. Pump power $P_\mathrm{in}=50$mW corresponds to $f\approx 4.11$. It was checked that qualitatively similar results can be obtained for other parameters of the microresonator, in particular for integrated SiN microrings \cite{Brasch14}.

In the simulations coupled mode equations for $501$ modes were numerically propagated in time using Runge-Kutta integrator with weak noise-like initial conditions. Nonlinear terms were calculated using the FFT acceleration proposed in \cite{Hansson14}. We checked that the results do not change with the increase in the number of modes. Simulations were performed using a in-house and open-source CombGUI toolbox \cite{Lihachev14} developed in MATLAB that allows numerical simulation of Kerr frequency comb.

The pump frequency was adiabatically scanned with time ($\tfrac{d\zeta_0}{d\tau}=0.1$) from the effective blue-detuned regime $\omega_p$ through the zero detuning frequency into the effectively red-detuned regime: $\zeta_0\in[-20;20]$. For analysis we calculated average intracavity intensity $U=\sum_\mu |a_\mu|^2$ as a function of normalized detuning $\zeta_0$ for different values of the pump mode shift $\Delta$.

\begin{figure}[htbp]
\centerline{\includegraphics[width=12cm]{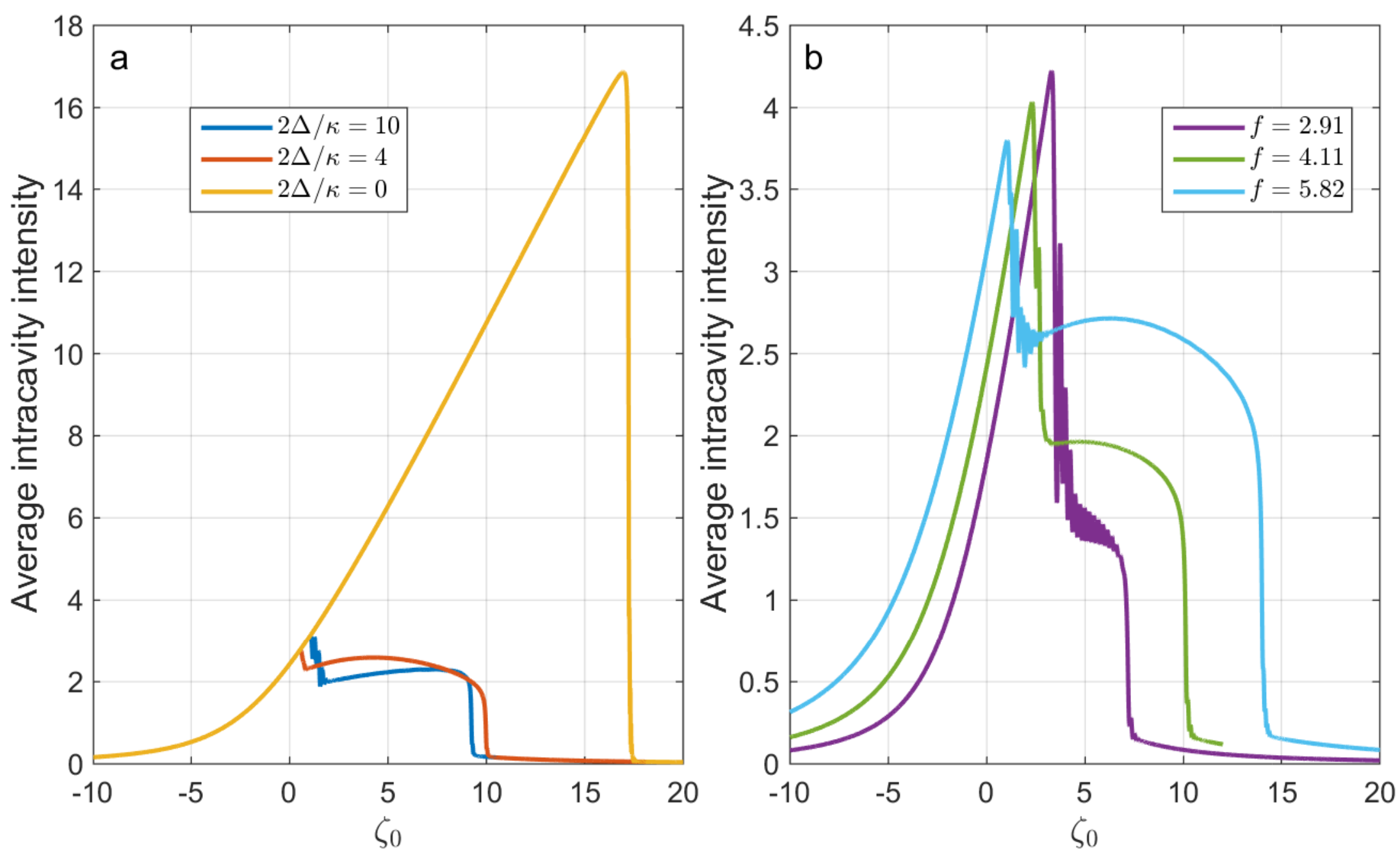}}
\caption{Intracavity average intensity $U$ vs. detuning $\zeta_0$ in normal GVD microresonators. Left: Same input power, different perturbations for $2\Delta/\kappa=0$ (no comb) and $2\Delta/\kappa=4,10$. $P_\mathrm{in}=50$mW. Right:  $2\Delta/\kappa=8$,  $P_\mathrm{in}=25$, 50 and 100 mW.}
\label{triangles}
\end{figure}

At $\Delta\leq 0$ conventional triangular nonlinear resonance corresponding to c.w. solutions for which just the pump mode is observed and no sideband modes are excited. However, at large enough positive values of eigenfrequency shift internal intensity vs. detuning  dependence deviates from the expected triangular shape and characteristic step-like dependence appears. Analogous discrete steps were identified as signatures of the soliton regime in microresonators with anomalous GVD \cite{Herr14}. However, in case of normal GVD instead of multiple steps with sharp edges as observed in anomalous GVD, more smooth and rounded, usually single shelf is formed (Fig.\ref{triangles}a). We found also that with the increase of the pump power, characteristic round shelf in the resonance curve becomes more pronounced and shifts to larger values of $\zeta_0$ (Fig.\ref{triangles}b).

\begin{figure}[htbp]
\centerline{\includegraphics[width=12cm]{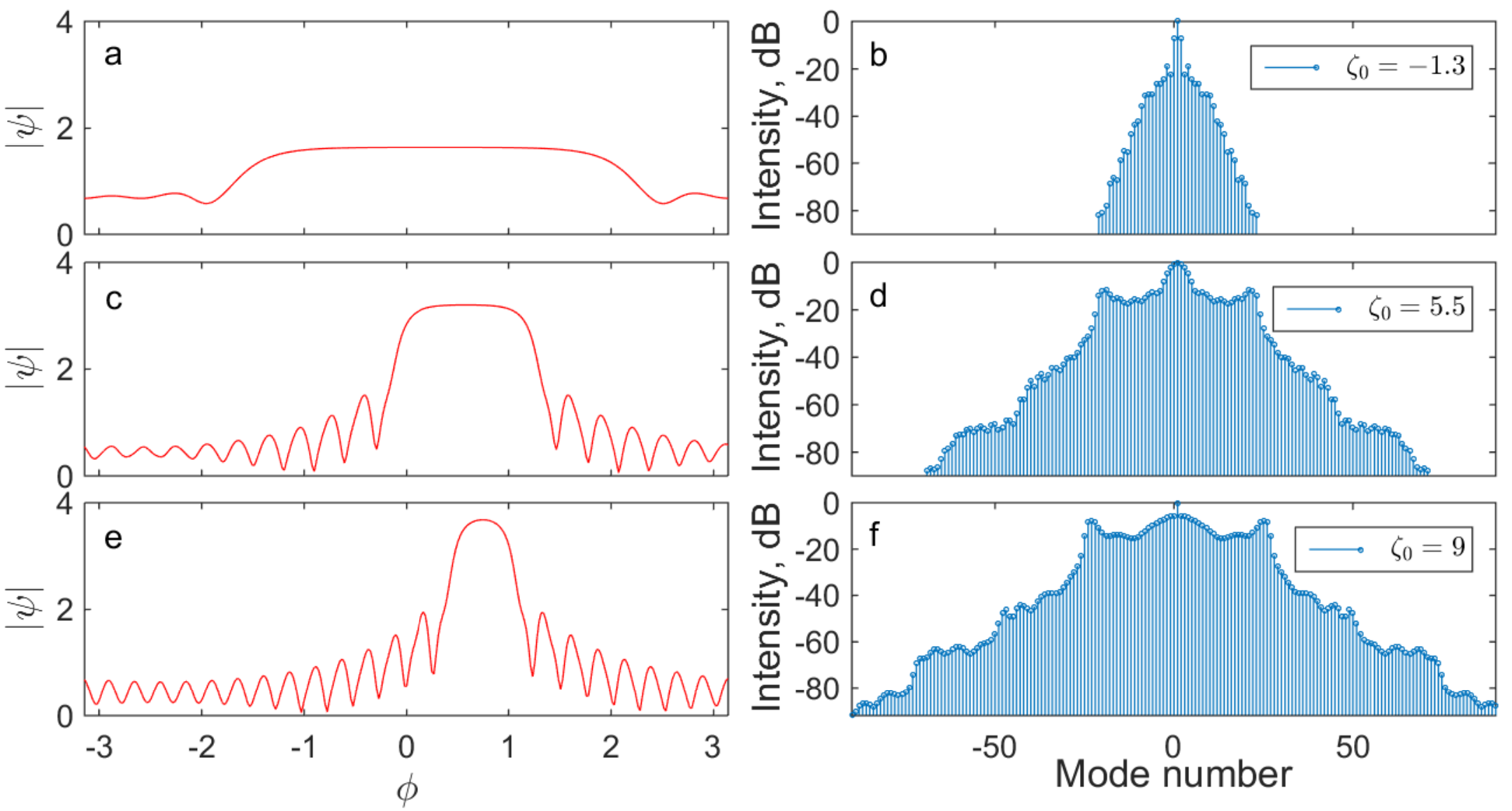}}
\caption{(Left: a, c, e) Numerically simulated shapes of optical platicons in normal GVD microresonators for different values of laser detunings. (Right: b, d, f) Corresponding optical spectra for the same set of parameters ($2\Delta/\kappa=4$, $P_\mathrm{in}=50$mW).}
\label{platicons}
\end{figure}

If the frequency scan is halted in the vicinity of the discrete step one observes steady-state regimes (see Fig.\ref{platicons}). The waveform was calculated as: $\psi(\phi)=\sum_\mu a_\mu \exp(i\mu\phi)$, where $\phi$ is the azimuthal angle (Fig.\ref{platicons}a). Surprisingly, the generated waveform looks not like dark solitons but more like bright flat-topped pulses (platicons). In fact for periodic pulses the difference between bright and dark ones disappears. One may identify a bright platicon as a separating wall between two consecutive dark solitons following each other at a distance corresponding to the round-trip time. So that the train of widest (close to roundtrip time) bright platicons corresponds to the train of narrow dark solitons and vice versa. From the other point of view one platicon may be considered as two following opposite domain walls (kink and anti-kink). These platicons circulating in the microresonator look similar to demonstrated ``flaticons'' in optical fibers with normal GVD \cite{Varlot13}.  However, while flaticons in fibers exhibit stable self-similar expansion of their central region, platicons' profile remains invariant upon circulation in microresonators. Frequency spectra of platicons (see Fig.\ref{platicons}, right) are characterized by two pronounced wings. Similar spectra were observed experimentally in nonlinear magnesium fluoride whispering-gallery mode resonator with a normal group velocity dispersion \cite{Liang14}.
It is especially interesting that the duration of the platicons depends strongly on the pump detuning. When the soliton regime is reached one may control effectively the duration and, consequently, the power of generated pulses slowly tuning the pump frequency. Figure \ref{platicons} shows that while pump frequency decreases (corresponding detuning increases), the pulse duration also decreases. Consequently, the width of the corresponding frequency comb may also be tuned. Such effective duration tuning is possible only for sufficient value of $\Delta$.

For higher values of pump power one may observe a resonance curve with several steps, where each step corresponds to the defined number of platicons. However, structures consisting of several platicons are less stable and transform into single platicon upon propagation.

\begin{figure}[htbp]
\centerline{\includegraphics[width=12cm]{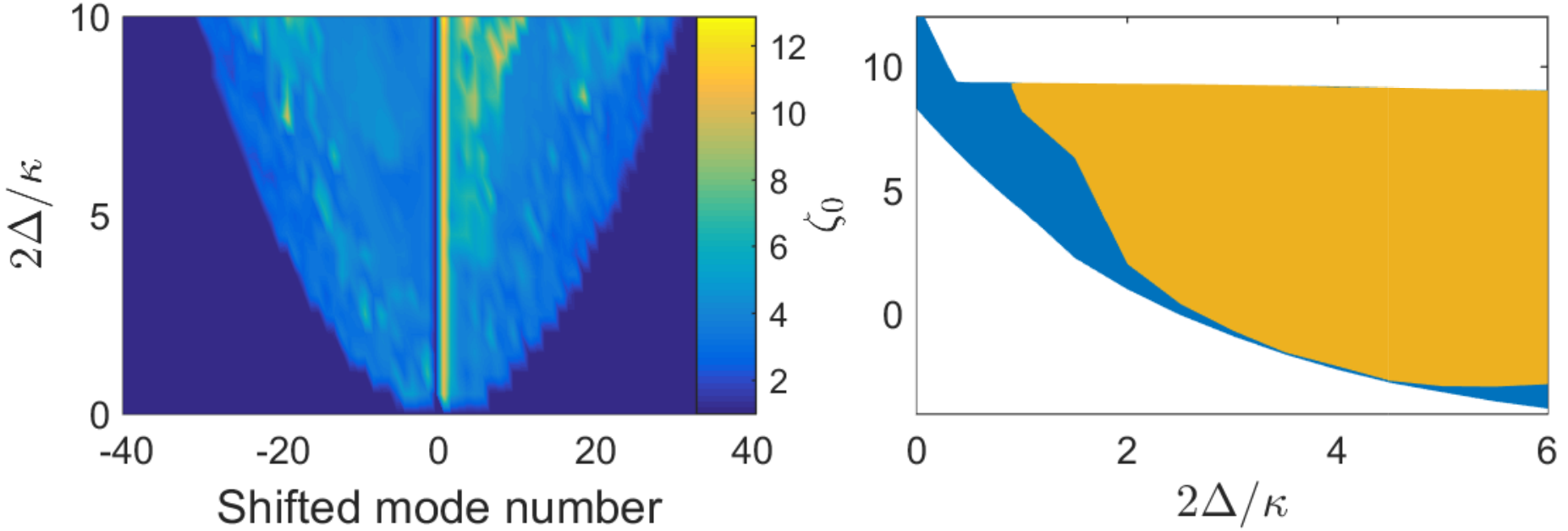}}
\caption{(Left) Numerical investigation of pulse formation in normal GVD microresonators. Color coded ratio of peak to average intensity in resonator, used as an indication of pulses, is mapped for different eigenmodes $\mu_s$ with eigenfrequencies shifted by $\Delta$. Central bright straight vertical line corresponds to $\mu_s=0$ with platicons analyzed in this paper. Right: Existence (blue) and soft excitation (yellow) domains of platicons.}
\label{existence}
\end{figure}

Through numerical simulation we found that the most efficient method to excite platicons is to shift the eigenfrequency of the pumped mode $\mu_s=0$ (see Fig.\ref{existence}, left). We used the same mapping of the peak to average intensity ratio for different parameters which was used to reveal bright solitons in anomalous GVD microresonators \cite{Herr14b}.

To calculate the existence domain of platicons we used both coupled mode approach and the model based on Lugiato-Lefever equation \cite{Godey14,Lugiato87,Chembo13} modified as described in \cite{Xue14} to include mode interaction. Results obtained by two methods were found to be in a good agreement.  It was revealed that platicons can exist even at $\Delta=0$  and in this case their existence domain coincides with the existence domain of dark solitons described in \cite{Godey14}. At small values of $\Delta$ several platicon solutions may exist at the same point inside the existence domain. However, dynamic soft excitation of platicons is possible only if $\Delta >\Delta_\mathrm{cr}$.
To confirm this, we found an excitation parameter domain where it is possible to generate platicons upon halting the pump frequency scan.
For example, for $f=4.11$ generation of platicons is possible if $2\Delta/\kappa>0.9$ (Fig.\ref{existence}, right).
We found also that soft excitation of platicons is possible without laser scanning at fixed pump frequency from noise-like input.

\begin{figure}[htbp]
\centerline{\includegraphics[width=12cm]{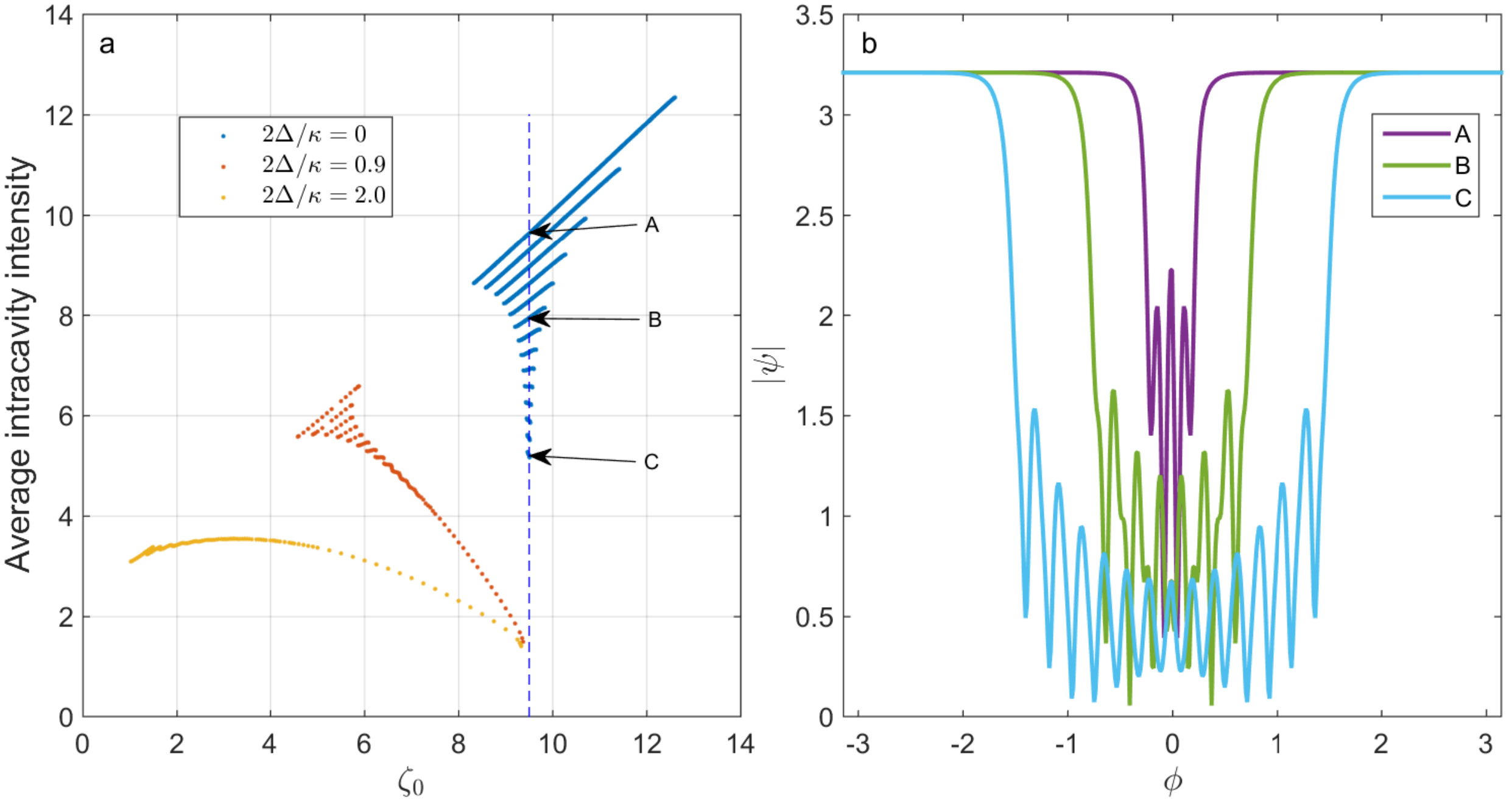}}
\caption{Transition of discrete spectrum of dark solitons into continuous platicons upon increased perturbation $\Delta$ of the pumped mode frequency at $P_\mathrm{in}=50$ mW (left). Several higher order dark dissipative solitons (right).}
\label{famdetune}
\end{figure}

An interesting question is the relation of platicons to dark solitons which are the solutions of the conserved Lugiato-Lefever equation without dissipation. We found that in unperturbed system at $\Delta =0$  several types of dark solitons may also exist in a narrow frequency domain (see Fig. \ref{famdetune}). Here average intracavity intensity was calculated as $U=\tfrac{1}{2\pi}\int_0^{2\pi} |\psi(\phi)|^2 d\phi$. Higher order solitons with larger number of oscillations and wider dip possess narrower existence domain (compare corresponding profiles at Fig. \ref{famdetune}). The existence domain of all soliton families remains inside the existence domain of the narrowest dark soliton (widest platicon). In this case frequency tuning may result in weak step-like pulse widening only. However, if a pump frequency shift is introduced, the discrete spectrum transforms significantly. Firstly, energy levels of different soliton families shift relative to each other and, secondly, discrete levels merge into one wide level that allows continuous tuning of pulse duration.
Note, that pronounced angled turn of the existence domain for $2\Delta/\kappa\approx 0.35$ depicted at Fig. \ref{existence} indicates transformation of the discrete energy spectrum into quasi-continuous one. Soft excitation of platicons from noise (which is not possible for dark solitons) is possible after this transformation. Another interesting feature of platicons is that while conventional dark solitons at $\Delta =0$ may be considered as robust intermediate states between the asymptotic levels of the two stable steady states, minimal and maximal amplitudes of the platicons do not coincide with the amplitudes of c.w. flat solutions.

\begin{figure}[htbp]
\centerline{\includegraphics[width=12cm]{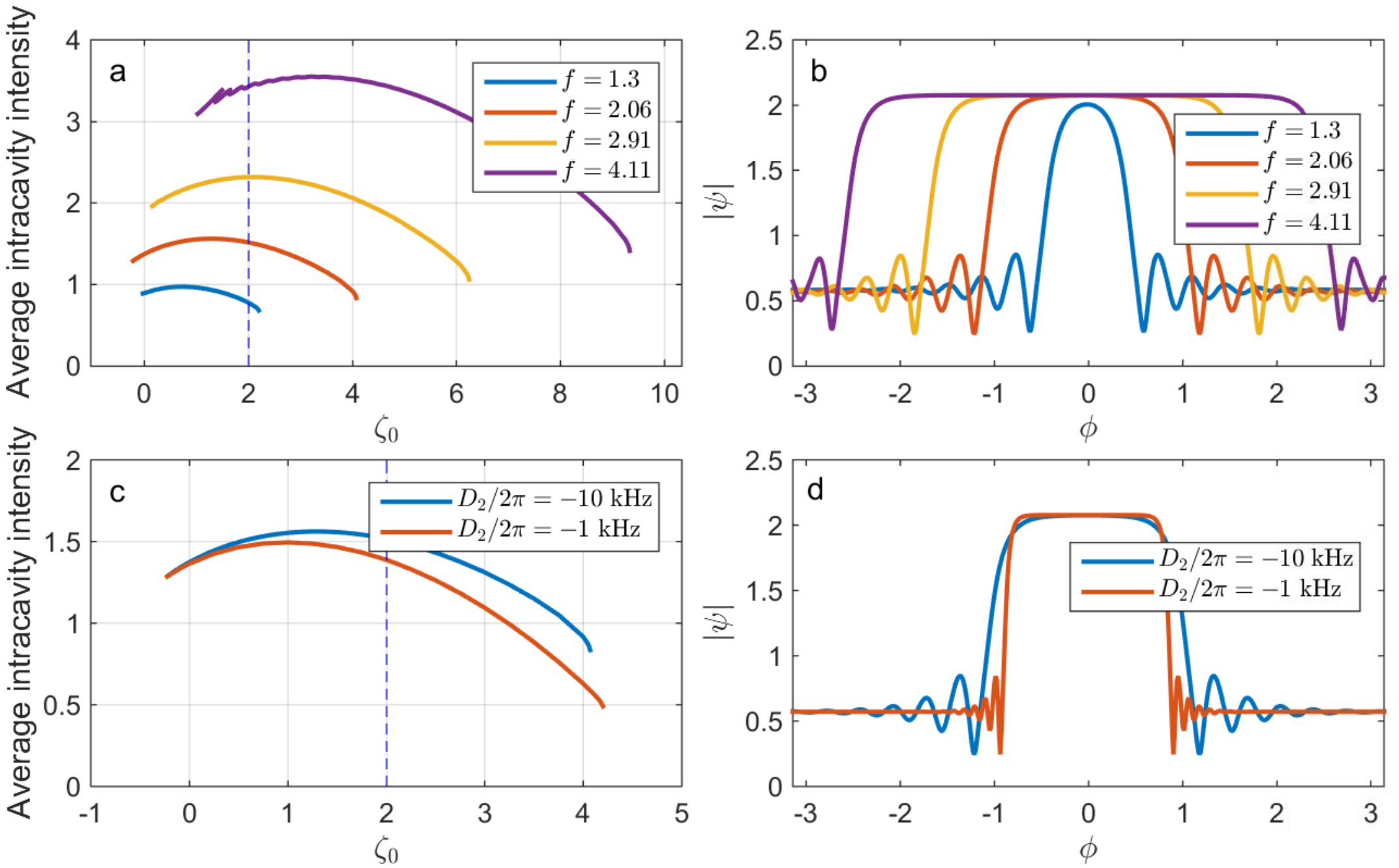}}
\caption{Families of platicons for different pump power (a,b). Families of platicons for different dispersion (c,d). In all cases $2\Delta/\kappa=2$.}
\label{famdisp}
\end{figure}

We also revealed that while the pump power decreases, the existence domain of platicons becomes narrower and shifts to smaller values of $\zeta_0$ (see Fig. \ref{famdisp}) so that platicon duration may also be controlled with pump power. Decreasing pump one may shift the platicon position inside the existence domain towards the upper boundary that corresponds to the narrowest possible platicons and, thus, generate shorter pulses (see Fig. \ref{famdisp}a,b). However, the domain weakly depends on the GVD value.
For example, we checked that at $2\Delta/\kappa=2$ and $f=2.06$ 10-fold decrease of dispersion coefficient $D_2$ results in more localized pulses with sharper profiles and shorter oscillating tails but practically does not affect their existence domain (see Fig. \ref{famdisp}c,d). In this way our results are applicable for a wide range of materials used for fabrication of microresonators with normal GVD.

Alternative methods of platicon generation may be proposed. It is known that injection locking pulls the resonant frequencies of the
laser and resonator to each other. In this way injection locking may be equivalent to the shift of the resonant eigenfrequency which is produced not by coupling between different families of modes but by coupling between the modes of the laser and the microresonator. We suppose that this effect resulted in observation of platicon spectra in \cite{Liang14}. One can also apply bi-harmonic pump using two lasers or utilizing amplitude/phase modulated pump. Let us consider the system where the first laser pumps the central mode with  $\mu=0$, the second laser pumps the neighboring mode with $\mu=1$. In this case it is important to have a frequency difference between two pump waves to be close to FSR. Under this condition soft generation of platicons is possible in a narrow domain of pump frequencies if relative power of the second pump wave exceeds the critical value. This critical value depends on the power of the first pump wave and increases with its growth. Outside the excitation domain of platicons low-contrast solutions are generated.

\begin{figure}[htbp]
\centerline{\includegraphics[width=12cm]{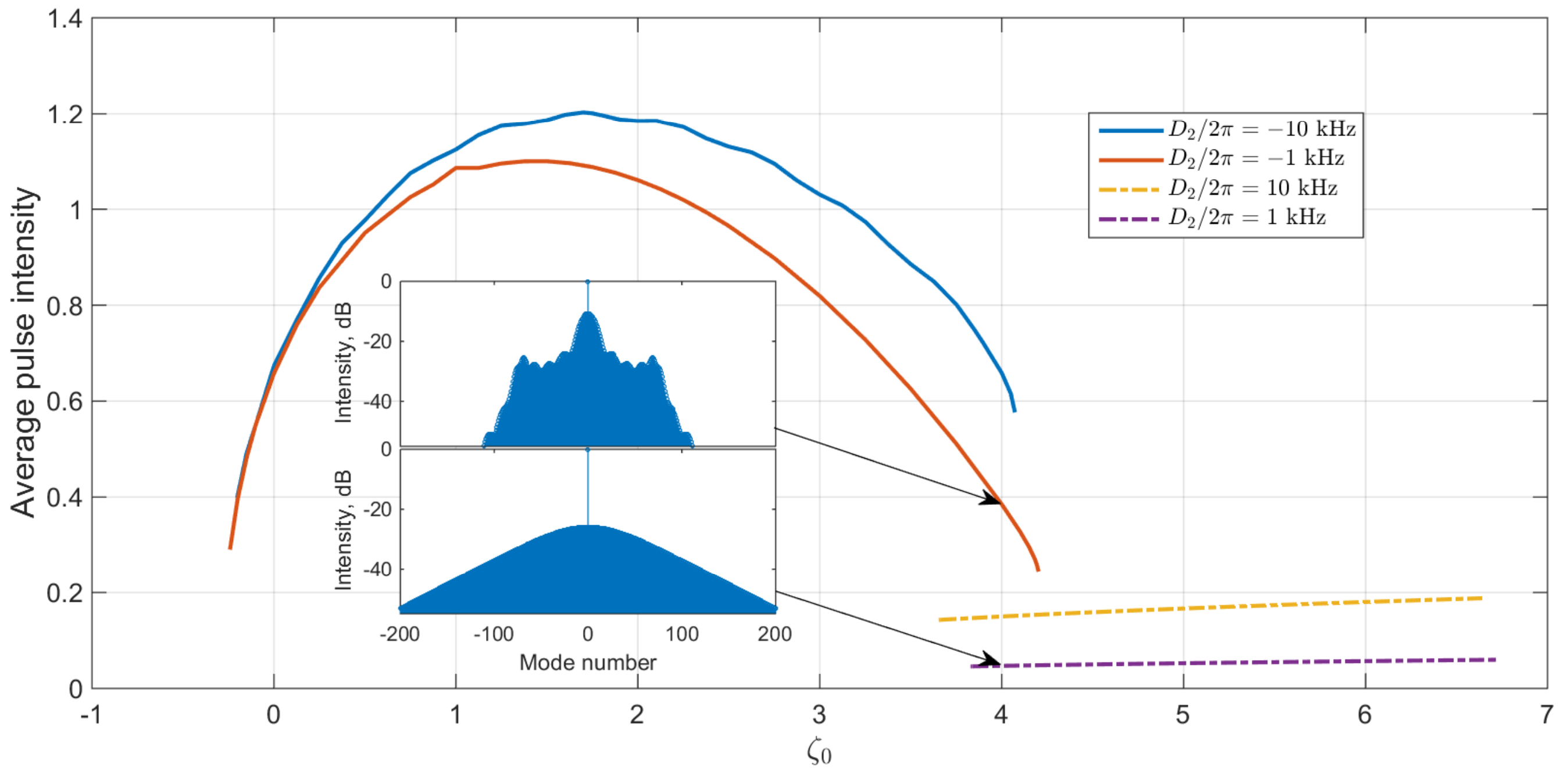}}
\caption{Comparison of average pulse intensity for the same absolute values of normal ($2\Delta/\kappa=2$, $P_\mathrm{in}=12.5$mW) and anomalous GVD combs. Insets show corresponding comb spectra.}
\label{efficiency}
\end{figure}

Normal dispersion microresonator based Kerr combs are interesting for future applications as they should allow obtaining coherent combs in visible and near infrared regions with normal dispersion. Moreover we found that these combs do not have the problem of low efficiency conversion of c.w. pump to the comb power degrading with an increased comb width \cite{Bao14}. As Fig.\ref{efficiency} demonstrates, the conversion efficiency for the platicon combs is significantly better for the same absolute value of dispersion and slowly scales with comb width. The best conversion is achieved when platicon width is close to
half of the roundtrip time. As insets of Fig.\ref{efficiency} show, first sidebands adjacent to the pump for the normal GVD comb are significantly stronger than in case of anomalous GVD, which may be advantageous for comb based photonic RF oscillators.

\section{Summary}
In conclusion, we found out that in a microresonator with normal GVD transformation of c.w. laser beam into a train of flat-top bright pulses, platicons, may occur in case of a small perturbation of the mode spectrum or with modulated pump. For these platicons one may control the duration of generated pulses varying slowly the pump frequency. The conversion efficiency of the c.w. pump to the frequency comb corresponding to platicons is significantly better than in case of anomalous dispersion bright soliton based combs.

\section*{Acknowledgements}
{This work was supported by the RFBR grant 13-02-00271.}


\begin{thebibliography}{10}
\newcommand{\enquote}[1]{``#1''}

\bibitem{DelHaye07}
P.~Del'Haye, A.~Schliesser, O.~Arcizet, T.~Wilken, R.~Holzwarth, and T.~J.
  Kippenberg, \enquote{Optical frequency comb generation from a monolithic microresonator,} Nature \textbf{450}, 1214--1217 (2007).

\bibitem{Savchenkov08}
A.~A. Savchenkov, A.~B. Matsko, V.~S. Ilchenko, I.~Solomatine, D.~Seidel, and
  L.~Maleki, \enquote{Tunable optical frequency comb with a crystalline whispering gallery mode resonator,} Phys Rev Lett \textbf{101}, 093902
  (2008).

\bibitem{Levy10}
J.~S. Levy, A.~Gondarenko, M.~A. Foster, A.~C. Turner-Foster, A.~L. Gaeta, and
  M.~Lipson, \enquote{CMOS-compatible multiple-wavelength oscillator for on-chip optical interconnects,} Nature Photonics \textbf{4}, 37--40 (2010).

\bibitem{Razzari10}
L.~Razzari, D.~Duchesne, M.~Ferrera, R.~Morandotti, S.~Chu, B.~E. Little, and
  D.~J. Moss, \enquote{CMOS-compatible integrated optical hyper-parametric oscillator,} Nature Photonics \textbf{4}, 41--45 (2010).

\bibitem{Kippenberg11}
T.~J. Kippenberg, R.~Holzwarth, and S.~A. Diddams,
  \enquote{Microresonator-based optical frequency combs,} Science \textbf{332},
  555--559 (2011).

\bibitem{Papp13}
S.~B. Papp, P.~Del'Haye, and S.~A. Diddams, \enquote{Mechanical control of a
  microrod-resonator optical frequency comb,} Physical Review X \textbf{3}, 7
  (2013).

\bibitem{Li12}
J.~Li, H.~Lee, T.~Chen, and K.~J. Vahala, \enquote{Low-pump-power,
  low-phase-noise, and microwave to millimeter-wave repetition rate operation
  in microcombs,} Physical Review Letters \textbf{109} (2012).

\bibitem{DelHaye08}
P.~Del'Haye, O.~Arcizet, A.~Schliesser, R.~Holzwarth, and T.~J. Kippenberg,
  \enquote{Full stabilization of a microresonator-based optical frequency
  comb,} Physical Review Letters \textbf{101} (2008).

\bibitem{DelHaye12}
P.~Del'Haye, S.~B. Papp, and S.~A. Diddams, \enquote{Hybrid electro-optically
  modulated microcombs,} Physical Review Letters \textbf{109} (2012).

\bibitem{Ferdous11}
F.~Ferdous, H.~X. Miao, D.~E. Leaird, K.~Srinivasan, J.~Wang, L.~Chen, L.~T.
  Varghese, and A.~M. Weiner, \enquote{Spectral line-by-line pulse shaping of
  on-chip microresonator frequency combs,} Nature Photonics \textbf{5},
  770--776 (2011).

\bibitem{Papp11}
S.~B. Papp and S.~A. Diddams, \enquote{Spectral and temporal characterization
  of a fused-quartz-microresonator optical frequency comb,} Physical Review A
  \textbf{84}, 7 (2011).

\bibitem{DelHaye11}
P.~Del'Haye, T.~Herr, E.~Gavartin, M.~L. Gorodetsky, R.~Holzwarth, and T.~J.
  Kippenberg, \enquote{Octave spanning tunable frequency comb from a
  microresonator,} Phys Rev Lett \textbf{107}, 063901 (2011).

\bibitem{Herr12}
T.~Herr, K.~Hartinger, J.~Riemensberger, C.~Y. Wang, E.~Gavartin, R.~Holzwarth,
  M.~L. Gorodetsky, and T.~J. Kippenberg, \enquote{Universal formation dynamics
  and noise of Kerr-frequency combs in microresonators,} Nature Photonics
  \textbf{6}, 480--487 (2012).

\bibitem{Herr14}
T.~Herr, V.~Brasch, J.~D. Jost, C.~Y. Wang, N.~M. Kondratiev, M.~L. Gorodetsky,
  and T.~J. Kippenberg, \enquote{Temporal solitons in optical microresonators,}
  Nat Photon \textbf{8}, 145--152 (2014).

\bibitem{Brasch14}
V.~Brasch, T.~Herr, M.~Geiselmann, G.~Lihachev, M.~H.~P. Pfeiffer, M.~L.
  Gorodetsky, and T.~J. Kippenberg, \enquote{Photonic chip based optical
  frequency comb using soliton induced Cherenkov radiation,}  (2014).
  \url{http://arxiv.org/abs/1410.8598}.

\bibitem{Jost14}
J.~D. Jost, T.~Herr, C.~Lecaplain, V.~Brasch, M.~H.~P. Pfeiffer, and T.~J.
  Kippenberg, \enquote{Microwave to optical link using an optical
  microresonator,}  (2014). \url{http://arxiv.org/abs/1411.1354}.

\bibitem{Matsko12}
A.~B. Matsko, A.~A. Savchenkov, and L.~Maleki, \enquote{Normal group-velocity
  dispersion Kerr frequency comb,} Opt Lett \textbf{37}, 43--5 (2012).

\bibitem{Hansson13}
T.~Hansson, D.~Modotto, and S.~Wabnitz, \enquote{Dynamics of the modulational
  instability in microresonator frequency combs,} Physical Review A
  \textbf{88}, 023819 (2013).

\bibitem{Godey14}
C.~Godey, I.~V. Balakireva, A.~Coillet, and Y.~K. Chembo, \enquote{Stability
  analysis of the spatiotemporal lugiato-lefever model for Kerr optical
  frequency combs in the anomalous and normal dispersion regimes,} Physical
  Review A \textbf{89}, 063814 (2014).

\bibitem{Coillet13}
A.~Coillet, I.~Balakireva, R.~Henriet, K.~Saleh, L.~Larger, J.~M. Dudley, C.~R.
  Menyuk, and Y.~K. Chembo, \enquote{Azimuthal Turing patterns, bright and dark
  cavity solitons in Kerr combs generated with whispering-gallery-mode
  resonators,} IEEE Photonics Journal \textbf{5}, 9 (2013).

\bibitem{Liang14}
W.~Liang, A.~A. Savchenkov, V.~S. Ilchenko, D.~Eliyahu, D.~Seidel, A.~B.
  Matsko, and L.~Maleki, \enquote{Generation of a coherent near-infrared Kerr
  frequency comb in a monolithic microresonator with normal GVD,} Optics
  Letters \textbf{39}, 2920--2923 (2014).

\bibitem{Liang14b}
  W.~Liang, D.~Eliyahu, V.~Ilchenko, A.~Savchenkov, A.~Matsko, and L.~Maleki,
  \enquote{All-optical micro-clock,} in \enquote{Frequency Control Symposium
  (FCS), 2014 IEEE International,} pp. 1--4.

\bibitem{Xue14}
X.~Xue, Y.~Xuan, Y.~Liu, P.~Wang, S.~Chen, J.~Wang, D.~E. Leaird, M.~Qi, and
  A.~M. Weiner, \enquote{Mode interaction aided self excitation of dark
  solitons and offset frequency tuning in microresonators constructed of normal
  dispersion waveguides,}  (2014). \url{http://arxiv.org/abs/1404.2865}.

\bibitem{Herr14b}
T.~Herr, V.~Brasch, J.~D. Jost, I.~Mirgorodskiy, G.~Lihachev, M.~L. Gorodetsky,
  and T.~J. Kippenberg, \enquote{Mode spectrum and temporal soliton formation
  in optical microresonators,} Physical Review Letters \textbf{113}, 123901
  (2014).

\bibitem{Savchenkov12}
A.~A. Savchenkov, A.~B. Matsko, W.~Liang, V.~S. Ilchenko, D.~Seidel, and
  L.~Maleki, \enquote{Kerr frequency comb generation in overmoded resonators,}
  Opt Express \textbf{20}, 27290--8 (2012).

\bibitem{Liu14}
Y.~Liu, Y.~Xuan, X.~Xue, P.-H. Wang, S.~Chen, A.~J. Metcalf, J.~Wang, D.~E.
  Leaird, M.~Qi, and A.~M. Weiner, \enquote{Investigation of mode coupling in
  normal-dispersion silicon nitride microresonators for Kerr frequency comb
  generation,} Optica \textbf{1}, 137--144 (2014).

\bibitem{Varlot13}
B.~Varlot, S.~Wabnitz, J.~Fatome, G.~Millot, and C.~Finot,
  \enquote{Experimental generation of optical flaticon pulses,} Optics Letters
  \textbf{38}, 3899--3902 (2013).

\bibitem{Chembo10}
Y.~K. Chembo and N.~Yu, \enquote{Modal expansion approach to
  optical-frequency-comb generation with monolithic whispering-gallery-mode
  resonators,} Physical Review A \textbf{82} (2010).

\bibitem{Hansson14}
T.~Hansson, D.~Modotto, and S.~Wabnitz, \enquote{On the numerical simulation of
  Kerr frequency combs using coupled mode equations,} Optics Communications
  \textbf{312}, 134--136 (2014).

\bibitem{Lihachev14}
G.~V. Lihachev, T.~Herr, T.~J. Kippenberg, and M.~L. Gorodetsky, \enquote{Tool
  for simulation of Kerr frequency combs,} in \enquote{Microresonator Frequency
  Combs and Applications, Ascona, Switzerland,} (code:
  \url{http://www.rqc.ru/groups/gorodetsky/combgui.htm}).

\bibitem{Lugiato87}
L.~A. Lugiato and R.~Lefever, \enquote{Spatial dissipative structures in
  passive optical-systems,} Physical Review Letters \textbf{58}, 2209--2211
  (1987).

\bibitem{Chembo13}
Y.~K. Chembo and C.~R. Menyuk, \enquote{Spatiotemporal lugiato-lefever
  formalism for Kerr-comb generation in whispering-gallery-mode resonators,}
  Physical Review A \textbf{87} (2013).

\bibitem{Bao14}
C.~Bao, L.~Zhang, A.~Matsko, Y.~Yan, Z.~Zhao, G.~Xie, A.~M. Agarwal, L.~C.
  Kimerling, J.~Michel, L.~Maleki, and A.~E. Willner, \enquote{Nonlinear
  conversion efficiency in Kerr frequency comb generation,} Opt. Lett.
  \textbf{39}, 6126--6129 (2014).

\end{thebibliography}
\end{document}